\documentclass[10pt,aps,twocolumn, prl, footinbib]{revtex4-1}
\usepackage{graphicx}
\usepackage{psfrag}
\usepackage{amsmath,amssymb}
\usepackage{colordvi}
\usepackage{color}

\renewcommand{\a}{\alpha}
\renewcommand{\b}{\beta}

\newcommand{\G}{\Gamma}
\renewcommand{\d}{\delta}

\newcommand{\s}{\sigma}

\renewcommand{\th}{\theta}

\renewcommand{\o}{\omega}

\renewcommand{\dag}{\dagger}
\newcommand{\lb}{\label}
\newcommand{\nn}{\nonumber}

\newcommand{\BK}[1]{\left[#1\right]}

\newcommand{\bra}[1]{\left\langle{#1}\right|}
\newcommand{\ket}[1]{\left|{#1}\right\rangle}
\newcommand{\lr}[1]{\left\langle#1\right\rangle}

\newcommand{\be}{\begin{equation}}
\newcommand{\ee}{\end{equation}}
\newcommand{\ba}{\begin{eqnarray}}
\newcommand{\ea}{\end{eqnarray}}
\newcommand{\m}[4]{\left[\begin{array}{cc} #1&#2\\#3&#4 \end{array}\right]}

\renewcommand{\vr}[2]{\Big[\begin{array}{c} #1\\#2 \end{array}\Big]}

\begin{document}

\title{Quantum Phases and Collective Excitations in Bose-Hubbard Models with Staggered Magnetic Flux}
\author{Juan Yao}
\author{Shizhong Zhang}
\affiliation{Department of Physics and Centre of Theoretical and Computational Physics, The University of Hong Kong, Hong Kong, China}
\date{\today}
\begin{abstract}
We study the quantum phases of a Bose-Hubbard model with staggered magnetic flux in two dimensions, as has been realized recently [Aidelsburger {\it et al.}, PRL, {\bf 107}, 255301 (2011)]. Within mean field theory, we show how the structure of the condensates evolves from weak to strong coupling limit, exhibiting a tricritical point at the Mott-superfluid transition. Non-trivial topological structures (Dirac points) in the quasi-particle (hole) excitations in the Mott state are found within random phase approximation and we discuss how interaction modifies their structures. Excitation gap in the Mott state closes at different ${\bf k}$ points when approaching the superfluid states, which is consistent with the findings of mean field theory.
\end{abstract}
\maketitle

The possibility of achieving quantum Hall~\cite{qhe} and other topological states~\cite{QiZhang,HasanKane} in cold atom systems has been greatly enhanced recently with the realization of synthetic gauge fields in free-space~\cite{Lin2009a,Lin2009b,Lin2012,Cheuk2012,Wang2012,Zhang2012} and synthetic magnetic flux in optical lattices~\cite{Struck2011,Struck2012,Aidelsburger2011,Aidelsburger2013,Miyake2013}. In the latter case, the flux per plaquette can approach the quantum limit, large enough to realize the Harper-Hafstadter Hamiltonian~\cite{Harper1955,Hofstadter1976} that hosts fractal energy spectrum (Hofstadter's butterfly).  With rational flux per plaquette, the lowest sub-band is topologically non-trivial and can give rise to quantized Hall conductance~\cite{qhe}. Such topological band structures can be readily explored with non-interacting fermionic atoms. 

The interaction effects on topological states are in general less well understood and in this regard, the cold atom realization offers an ideal platform for addressing this issue.  In optical lattices, interaction strength can be tuned by adjusting the lattice depth and if necessary, with Feshbach resonance. Furthermore, cold atom systems allow the study of bosonic variant and open experimental avenue for investigating bosonic topological states that has attracted much theoretical attention recently~\cite{Lu2012,Chen2013,Senthil2013,Metlitski2013,Xu2013}. In the weak coupling limit, Bose condensation in a uniform flux has been analyzed with Bogoliubov theory~\cite{Powell2010}. Quantum phases of bosons in a staggered flux lattice has been investigated as well~\cite{Lim2008}. 

In this Letter, we study the interaction effects on the quantum phases of bosonic $^{87}$Rb atoms in an optical lattice with magnetic flux that is staggered only along one direction, as was realized recently in Ref.~\cite{Aidelsburger2011}. We show how two types of superfluid states in the weak coupling limit evolve into the Mott insulating regime through a tricritical point. Collective excitations in the Mott regime are studied in detail, and give further evidence for existence of tricritical point. Multiple Dirac points are found in the collective excitation and we investigate the effects of interactions on the Dirac points and show that while their dispersion depends strongly on interactions, their positions in the Brillouin zone remain intact in the Mott regime. 
\begin{center}
\begin{figure}[t]
\includegraphics[width=0.45 \textwidth]{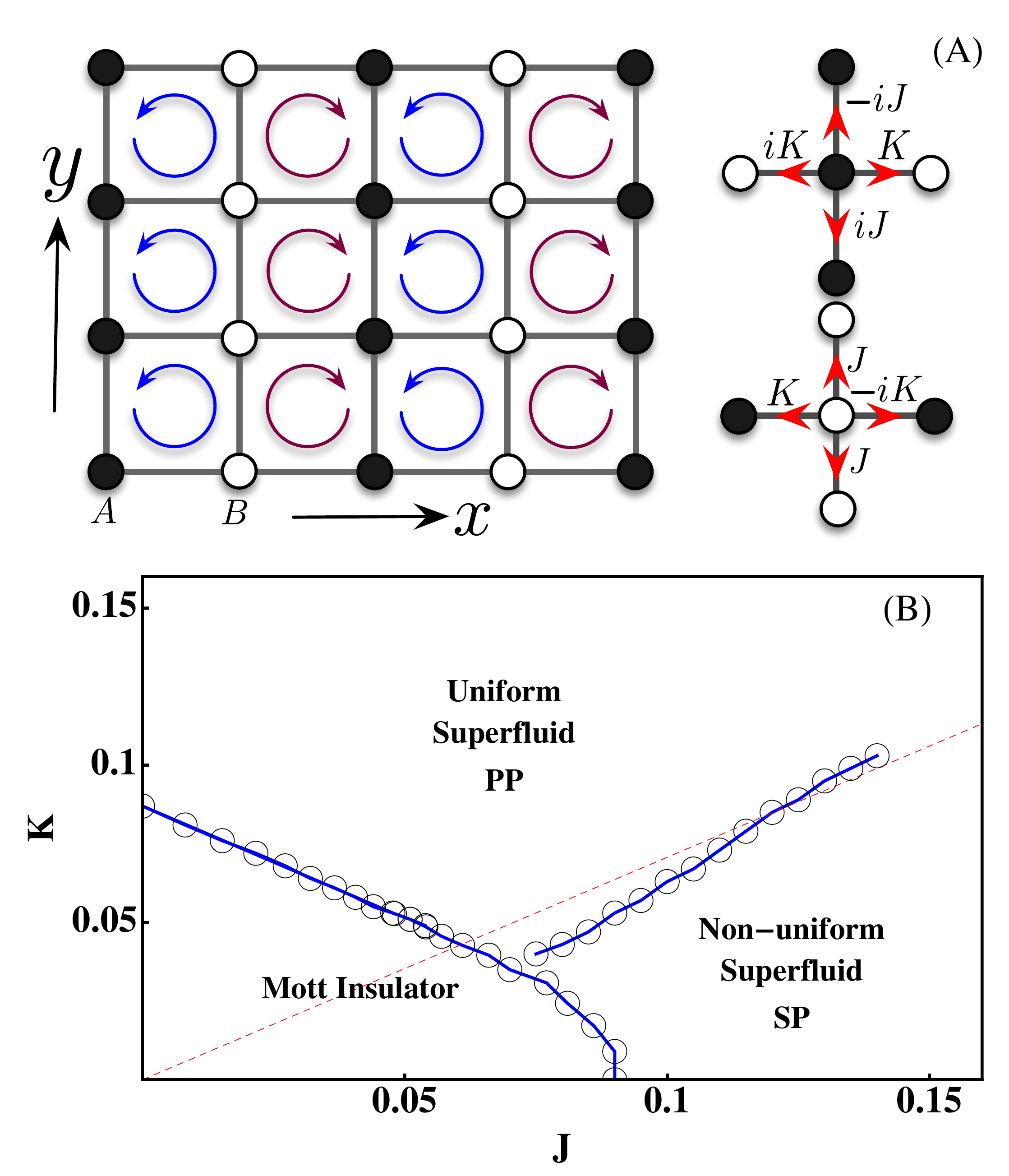}
\caption{(Color online). (A) Schematics of the two-dimensional square optical lattice with synthetic magnetic flux of magnitude $\frac{\pi}{2}$.  The flux is periodic along $\hat{y}$-direction, but staggered along $x$-direction. This gives rise to a unit cell with two non-equivalent lattice sites $A$ (filled circle) and $B$ (unfilled circle). The hopping amplitudes from $A$ and $B$ sites to their nearest neighbors are shown schematically. (B) Phase diagram based on the inhomogeneous mean field theory for average filling factor $n_{\rm A}+n_{\rm B}=2$. Three phases are found within the inhomogeneous mean field theory with cluster of dimension $8\times 8$. Besides the Mott insultating state, two superfluid states (plane wave phase (PP) and stripe phase (SP)) are found and they terminate at a tricritical point. The dashed red line gives the boundary between PP and SP for non-interacting gas.}
\label{fig:mft}
\end{figure}
\end{center}

{\em Single particle properties}. In the experiment~\cite{Aidelsburger2011}, magnetic flux per plaquette with a magnitude $\frac{\pi}{2}$ which is periodic along $\hat{y}$-direction, but staggered along $\hat{x}$-direction, is realized with laser assisted hopping in a superlattice; see Fig.\ref{fig:mft} (A). The magnitude of the hopping amplitude along the $\hat{x}$ and $\hat{y}$ directions are given by $K$ and $J$, respectively. Due to the presence of magnetic flux, the hopping Hamiltonian acquires Peierls phases and takes the form $H_{\rm hop}=-\sum_{\bf R}(Ke^{\pm i \d{\bf k}\cdot{\bf R}}c^\dag_{\bf R}c_{{\bf R}+a\hat{x}}+Jc^\dag_{\bf R}c_{{\bf R}+a\hat{y}}+{\rm H.c.})$, where ${\bf R}=m a\hat{x}+n a \hat{y}$ are lattice sites and $a$ is the lattice constant and will be set to unit in the following. $\pm$ sign in the Peierls phases refers to the even and odd sites along $\hat{x}$-direction. $m$ and $n$ are integers. By a simple gauge transformation for {\em only} the even sites along $\hat{x}$-direction, $c_{2m\hat{x}+n\hat{y}}\to e^{i2m\d k_x+in\d k_y}a_{mn}$, and rename the odd sites $a_{(2m+1)\hat{x}+n\hat{y}}\to b_{mn}$,  the space-dependent phase factors can be removed and one obtains a Hamiltonian with a unit cell that consists two non-equivalent sites~\cite{Aidelsburger2011} (see Fig.\ref{fig:mft} (A)), which we shall label as $A$-site and $B$-site ($m$ now labels the unit cell along $\hat{x}$-direction). In terms of these new operators, the single particle Hamiltonian takes the form
\begin{align}\nn
&H_{\rm hop} =-\sum_{m,n}(Ka_{m,n}^\dag b_{m,n}+Je^{i\d k_y}a_{m,n}^\dag a_{m,n+1}+{\rm H.c.})\\
&-\sum_{m,n}(Ke^{i\d k_x}b_{m,n}^\dag a_{m+1,n}+Jb_{m,n}^\dag b_{m,n+1}+{\rm H.c.}).
\label{eq:ham}
\end{align}
In momentum space, $H_{\rm hop}=-\sum_{\bf k}\psi^\dag_{\bf k}H({\bf k})\psi_{\bf k}$, where $H({\bf k})=H_0({\bf k})\boldsymbol{I}+{\bf H}({\bf k})\cdot\boldsymbol{\s}$, with $H_0=J[\cos(k_y+\d k_y)+\cos(k_y)]$, $H_x({\bf k})=K[\cos k_x+\cos(k_x+\d k_x)]$, $H_y({\bf k})=K[\sin(k_x+\d k_x)-\sin k_x]$ and $H_z({\bf k})=J[\cos(k_y+\d k_y)-\cos k_y]$. $\boldsymbol{\s}=(\s_x,\s_y,\s_z)$ are the Pauli matrices and $\psi^\dag_{\bf k}=(a^\dag_{\bf k},b^\dag_{\bf k})$. $\boldsymbol{I}$ is the two-by-two identity matrix. The $A,B$ sub-lattice constitutes a pseudo-spin half degree of freedom. The single particle spectrum constitutes two branches and are given by $E_\pm({\bf k})=H_0({\bf k})\pm \left|{\bf H}({\bf k})\right|$ with the corresponding pseudo-spin points either along or opposite the direction of ${\bf H}({\bf k})$. In the following, we set $\d k_x=\d k_y=\frac{\pi}{2}$, as was the case in Ref.\cite{Aidelsburger2011}. 

The original Hamiltonian obeys the combined symmetry operations of time reversal $\mathcal{T}$ and spatial translation along $\hat{x}$-direction by one unit of lattice constant, $T_{\hat{x}}$, namely $T_{\hat{x}}^{-1}\mathcal{T}^{-1}H_{\rm hop}\mathcal{T}T_{\hat{x}}=H_{\rm hop}$. With the transformed Hamiltonian Eq.(\ref{eq:ham}), the spectrum exhibits the symmetry $E_-({\bf k})=-E_+({\bf k}+\pi\hat{y})$. The lowest energy states depend on the ratio of $J/K$~\cite{Aidelsburger2011}. For $J/K<\sqrt{2}$, the ground state is at $(q_x,q_y)=(-\frac{\pi}{4},-\frac{\pi}{4})$; while for $J/K>\sqrt{2}$, there are two degenerate minimum at $(-\frac{\pi}{4},-\frac{\pi}{2}+\frac{r}{2})$ and $(-\frac{\pi}{4},-\frac{r}{2})$, where $r=\arcsin[2(K/J)^2]$. There are also two non-degenerate Dirac points at $\frac{1}{2}(\pi-\d k_x,-\d k_y)$ and $\frac{1}{2}(\pi-\d k_x,2\pi-\d k_y)$. By changing the angle between the two laser Raman beams, one can move the Dirac points around in the first Brillouin zone $[-\frac{\pi}{2},\frac{\pi}{2}]_{\hat{x}}\otimes[-\pi,\pi]_{\hat{y}}$. 

{\em Mean-field phase diagram.} The interaction between bosons can be taken to occur only within the same lattice site and assumes the simple form (within a unit cell)
\be
H_{\rm int}=\frac{U}{2}(n_{\rm A}(n_{\rm A}-1)+n_{\rm B}(n_{\rm B}-1)),
\ee
where $U>0$ is the onsite repulsion. In the strong coupling limit, $U\gg J,K$, the system enters the Mott insulating regime with one boson per site and a finite excitation gap. Within mean field theory, the ground state wave function takes the form $\ket{\Psi}=\prod_{m,n}a^\dag_{m,n}b^\dag_{m,n}\ket{\rm vac}$. As a result, there is no superfluid order: $\varphi_{\rm A}(m,n)=\varphi_{\rm B}(m,n)=0$, where $\varphi_{\rm A}(m,n)=\lr{a_{m,n}}$ and $\varphi_{\rm B}(m,n)=\lr{b_{m,n}}$ are the order parameters. Furthermore, the density is uniform. This is, however, no longer the case when the system enters into the superfluid states. In that case, it is possible for system to develop both the superfluid and the density order and this is indeed what we find within mean field theory.

In the weak coupling limit $J,K\gg U$, for $J/K<\sqrt{2}$, there is only single ground state and the condensate wave function can be written as 
\begin{equation}\lb{pp}
\vr{\varphi_{\rm A}}{\varphi_{\rm B}}_{\rm PP}=\sqrt{n}\frac{1}{\sqrt{2}}\vr{1}{e^{i\frac{\pi}{4}}}e^{-i\frac{1}{4}\pi (2m+n)},
\end{equation}
where $n$ is the average number per unit cell. The density is uniform and the phase of the condensate modulates along $\hat{x}$ and $\hat{y}$ direction with period of $8$ lattice sites. We label this as plane wave phase (PP). On the other hand, when $J/K>\sqrt{2}$, there are two degenerate minima and in the presence of interaction, a general ansatz for the ground state wave function can be written as a superposition of two spinor wave functions at momentum $(-\frac{\pi}{4},-\frac{\pi}{4}+q)$ and $(-\frac{\pi}{4},-\frac{\pi}{4}-q)$, lying along $\hat{y}$-direction, symmetric with respect to the point $(-\frac{\pi}{4},-\frac{\pi}{4})$~\cite{Yun2012}
\begin{align}\lb{sp}
\vr{\varphi_{\rm A}}{\varphi_{\rm B}}_{\rm SP}=&\sqrt{n}\Big[C_1\vr{\sin\frac{\th}{2}}{\cos\frac{\th}{2}e^{i\frac{\pi}{4}}}e^{-i\frac{1}{4}\pi (2m+n)}e^{iq n}\\\nn
&+C_2\vr{\cos\frac{\th}{2}e^{i\frac{3\pi}{4}}}{\sin\frac{\th}{2}}e^{-i\frac{1}{4}\pi (2m+n)}e^{-iqn}\Big],
\end{align}
where $q$ is a variational parameter that should be determined, together with $C_1$ and $C_2$, by minimizing the mean field energy $E(q,C_1,C_2)\equiv \lr{H_{\rm hop}+H_{\rm int}}$. We note that if $q=0$, then ansatz Eq.(\ref{sp}) reduces to Eq.(\ref{pp}). For later convenience, we set $|C_1|=\cos\a$ and $|C_2|=\sin\a$ anticipating the relative phase between $C_1$ and $C_2$ is irrelevant for energy minimization. $\cos\th(q)=\frac{\sqrt{2}J\sin q}{\sqrt{2J^2\sin^2 q+4K^2}}$. The mean field energy is then given by
\begin{align}\lb{mfe}\nn
E(q)=&-n\sqrt{2}J\cos q-n\sqrt{2J^2\sin^2 q+4K^2}\\
&+\frac{1}{8}n^2\BK{3+\cos 4\a - \frac{2K^2(1+3\cos 4\a)}{4K^2+2J^2\sin^2 q}}.
\end{align}
We need to minimize Eq.(\ref{mfe}) for various values of $J$ and $K$. For $n=2$, it turns out that in general there are three possible phases: (i) plane wave phase with $q=0$ (Eq.(\ref{pp})); (ii) stripe phase with $q\neq 0$ with, however, either $C_1$ or $C_2$ equals to zero (Eq.(\ref{sp})) and (iii) stripe phase with $q\neq 0$ and $|C_1|=|C_2|=\frac{1}{\sqrt{2}}$ (Eq.(\ref{sp})). For relatively small values of $K$ and $J$ ($K<3.2$ and $J<1.5$), only two phases (i) and (iii) remain and they persist towards the Mott-superfluid transition, which is consistent with the inhomogeneous mean field theory to be discussed below. According to Eq.(\ref{sp}), the density modulates along $\hat{y}$-direction with form $\frac{n}{4}\sin\th\cos(2q n)$, while stays uniform along $\hat{x}$-direction. On the other hand, the phase of the condensate modulates with period of $8$ along $\hat{x}$-direction while varies in general non-commensurately along $\hat{y}$-direction.



To investigate how two types of condensate structures, (i) and (iii), discussed above evolve into the Mott state, we make use of the standard mean field theory and decouple the hopping term as $a_{m,n}^\dag b_{m,n}=\varphi_{\rm A}(m,n)^*b_{m,n}+a_{m,n}^\dag \varphi_{\rm B}(m,n)-\varphi_{\rm A}^*(m,n)\varphi_{\rm B}(m,n)+\widetilde{a}^\dag_{m,n}\widetilde{b}_{m,n}$, and the fluctuation term $\widetilde{a}^\dag_{m,n}\widetilde{b}_{m,n}\equiv (a^\dag_{m,n}-\varphi_{\rm A}^*(m,n))(b_{m,n}-\varphi_{\rm B}(m,n))$ is neglected. With similar decoupling scheme for other hopping terms, one obtains an effective single site Hamiltonian for the $A$-sub-lattice
\begin{align}\nn\lb{mfham}
&H_{\rm MF}^{(A)}=\frac{U}{2}(n_{\rm A}(n_{\rm A}-1))-[K\varphi_{\rm B}^*(m,n)+iK\varphi_{\rm B}^*(m-1,n)\\
&+iJ\varphi_{\rm A}^*(m,n-1)-iJ\varphi_{\rm A}^*(m,n+1)]a_{m,n}+{\rm H.c.}.
\end{align}
Similarly, one can write down $H_{\rm MF}^{(B)}$ for the $B$-sublattice. $H_{\rm MF}^{(A,B)}$ couples to its nearest neighbors through the mean fields $\varphi_{\rm A}(m,n)$ and $\varphi_{\rm B}(m,n)$ that are in general non-uniform in space and will be determined self-consistently.

The mean field phase diagram is shown in Fig.\ref{fig:mft} (B) for an average of one particle per site $n=2$ and one finds three phases. For small $K$ and $J$, the system is in the Mott insulating state, while depending on the ratio of $J/K$, the strong coupling superfluid state exhibits two different phases. The plane wave phase (PP), which occurs when the ratio $J/K$ is small, has uniform density while the phases modulate along $\hat{x}$ and $\hat{y}$ direction with period of $8$ lattice sites; for larger ratio $J/K$, stripe phase (SP) with density modulation along $\hat{y}$ occurs while the phases modulate along $\hat{x}$ with period of $8$ lattice sites. These features are all reminiscent of the weak coupling superfluid phase and has been checked for larger cluster sizes~\cite{com1}. The three phases meet at a tricritical point which, within mean field theory, is at $J=0.07$ and $K=0.035$. We shall defer a detailed study of the tricritical point later~\cite{Yao}. The fact that the Mott state makes transitions to two different superfluid states can also be identified from the excitation spectrum of the Mott state to which we now turn.



\begin{center}
\begin{figure}[t]
\includegraphics[width=0.46 \textwidth]{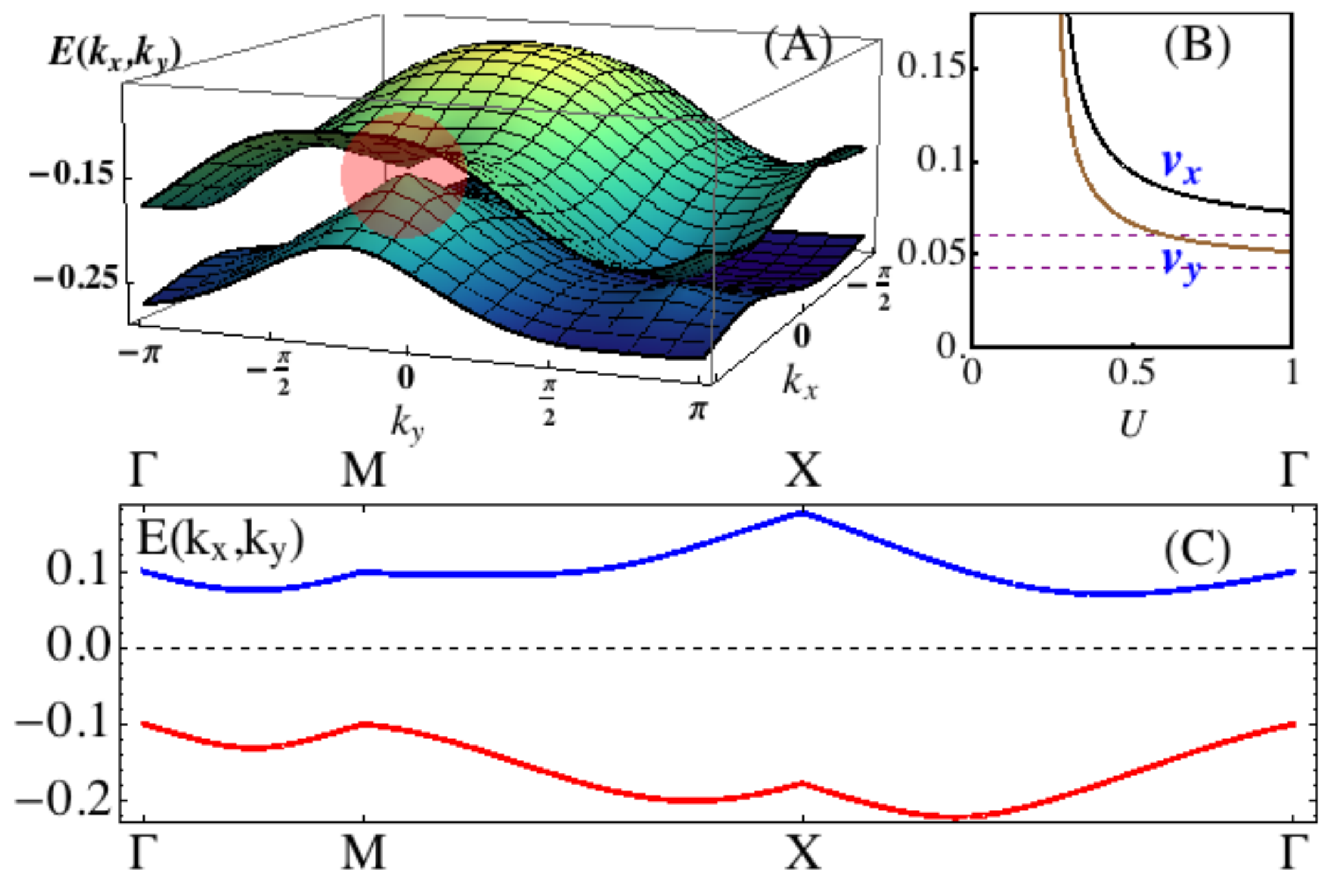}
\caption{(Color online). (A) Quasi-hole excitation for the two bands closest to zero energy. There are two Dirac points in the spectrum at ${\bf k}_1=(\frac{\pi}{4},\frac{3\pi}{4})$ and ${\bf k}_2=(\frac{\pi}{4},-\frac{\pi}{4})$. The shaded region shows the Dirac cone at ${\bf k}_2$. (B) The velocity of the quasi-hole along $k_x$ and $k_y$ direction for the Dirac point ${\bf k}_2$, as a function of $U$ in the Mott regime. The dashed line corresponds to the non-interacting value and has a ratio $\sqrt{2}$. (C) Quasi-particle and quasi-hole bands closest to zero energy. In the Mott regime, a finite gap exists between quasi-particle and quasi-hole band. The gap decreases as one approaches the superfluid phases. There is no particle-hole symmetry in the excitation spectrum. The following parameters are used for (A) and (C), $K=J=0.03$ and $U=1$. For (B), $K=J=0.03$. The average number of boson per site is set to one.}
\label{mott_excitation}
\end{figure}
\end{center}

\begin{center}
\begin{figure}[t]
\includegraphics[width=0.46 \textwidth]{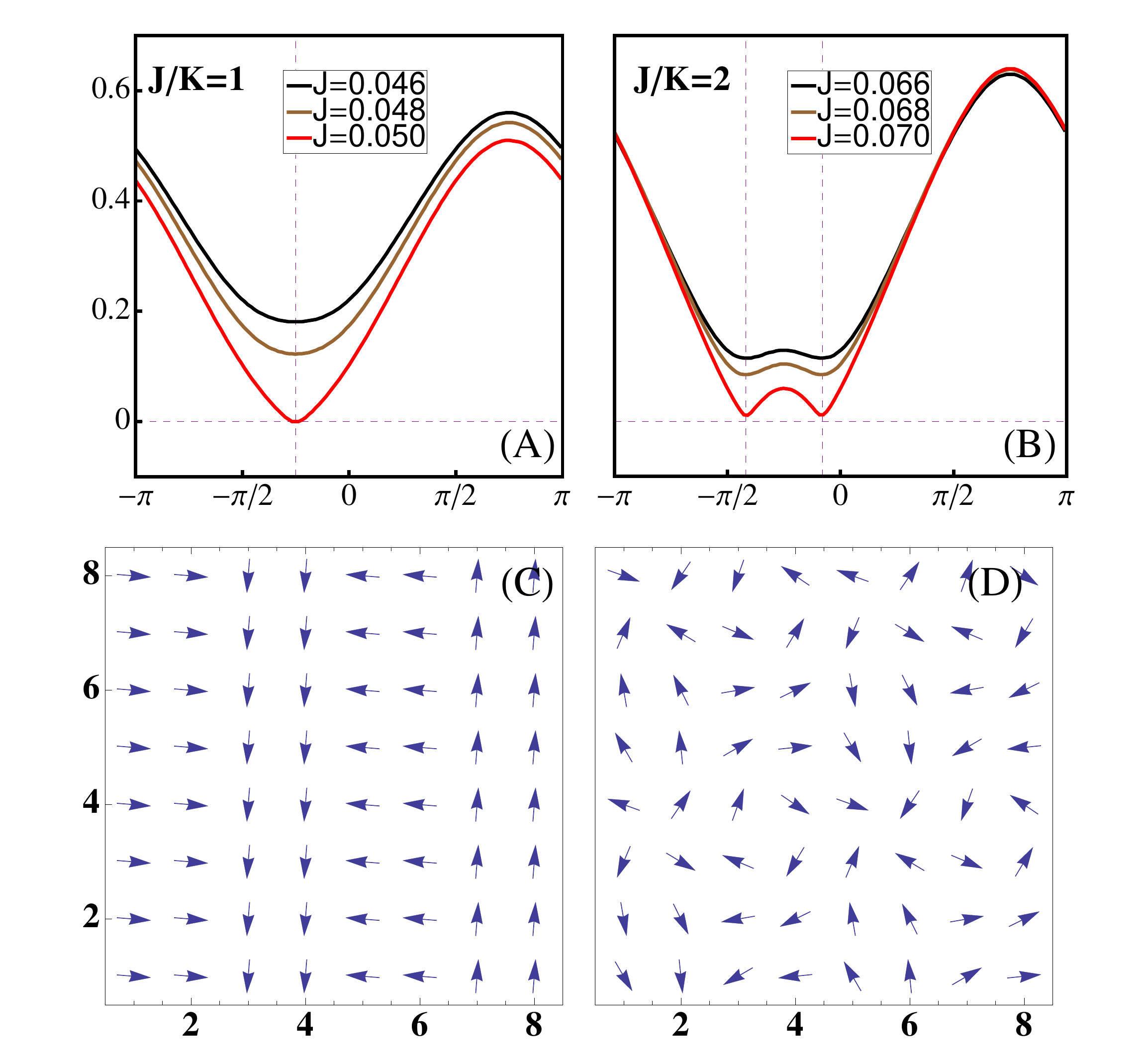}
\caption{(Color online). Softening of the excitation spectrums in Mott insulating state as a function of $k_y$, for $k_x=-\pi/4$, along two values of $J/K$, making transition to PP and SP. (A) $J/K=1$, the excitation spectrum goes to zero at $k_x=-\pi/4,k_y=-\pi/4$, consistent with the structure of the strong coupling superfluid state. (B) $J/K=2$, excitation spectrum goes to zero at two values of $k_y$, indicating the emergence of condensate with two momentum conponents. (C) and (D) shows the phases (directions of the arrows) of superfluid order parameters at each site for PP and SP, respectively. Overall $\pi/4$ modulation along $y$ has been removed to make comparsion between (C) and (D) clear. The interaction parameter $U$ is set to unity.}
\label{excitation}
\end{figure}
\end{center}

{\em Collective excitations.} Having established the mean field phase diagram, let us now discuss the collective excitations in the Mott insulating phase. Since density is uniform in the Mott regime (one particle per site), the unit cell turns out to be composed of $A$ and $B$ sites, as in the non-interacting case. Let us then define the local basis for the mean field Hamiltonian $H_{\rm MF}^{(\chi)}$ ($\chi$=A,B) as $\ket{\chi;\a}$, with eigen-energy $E^\chi_\a$. If we now define the standard basis operators $L^\chi_{\a\a'}=\ket{\chi,\a}\bra{\chi,\a'}$, where $\a,\a'$ label the eigenstates, then $H_{\rm MF}^{(\chi)}=\sum_\a E^\chi_\a L^\chi_{\a\a}$, diagonal in the basis $\ket{\chi,\a}$. On the other hand, the fluctuation terms that have been neglected in the mean field treatment, can now be written in terms of $L^\chi_{\a\a'}$: $\widetilde{a}^\dag_{m,n}\widetilde{b}_{m,n}=\sum_{\a\a';\b\b'}D^{+A}_{\a\a'}D^{-B}_{\b\b'}L^{A}_{\a\a'}L^{B}_{\b\b'}$ and $D^{\s\chi}_{\a\a'}=\bra{\chi,\a}\psi^\s_\chi-\langle\psi^\s_\chi\rangle\ket{\chi,\a'}$, where $\s=\pm$ in the subscript denotes creation and annihilation operators and $\psi^\s_A=a^\s$ and $\psi^\s_B=b^\s$. The full original Hamiltonian can now be written in terms of the operators $L_{\a\a'}^\chi$. 

To obtain the excitation spectrum, we define the single particle Green function $\boldsymbol{G}_{\chi\chi'}^{\s\s'}({\bf r},{\bf r}';t)=-i\langle\mathcal{T}\psi_\chi^\s({\bf r},t)\psi_{\chi'}^{\s'}({\bf r}',0)\rangle$, where $\mathcal{T}$ is the time-ordering operator. Writing $\boldsymbol{G}_{\chi\chi'}^{\s\s'}$ in terms of standard basis operators and making use of the random phase approximation~\cite{Sheshadri1993,Ohashi2006}, we find the following equation for the Green function $\boldsymbol{G}_{\chi\chi'}^{\s\s'}$ (after Fourier transforming to the frequency-momentum space)
\begin{align}\lb{eomG}
&\boldsymbol{G}_{\chi\chi'}^{\s\s'}({\bf q},\o)=\frac{1}{2\pi}\d_{\chi\chi'}\boldsymbol{B}^{\s\s'}_{\chi}(\o)-2\sum_i e^{i{\bf q}\cdot{\boldsymbol{\d}_i}}\times\\\nn
&\Big[\boldsymbol{B}_\chi^{\s-}(\o)t_{\boldsymbol{\d}_i}^\chi \boldsymbol{G}_{\chi_i\chi'}^{+\s'}({\bf q},\o)+\boldsymbol{B}_\chi^{\s+}(\o)t_{\boldsymbol{\d}_i}^{\chi*} \boldsymbol{G}_{\chi_i\chi'}^{-\s'}({\bf q},\o)\Big],
\end{align}
in which $\boldsymbol{\d}_i$'s are the four unit vectors which point outwards from a particular site. $t_{\boldsymbol{\d}_i}^\chi$ gives the hopping amplitude from a site with sub-lattice index $\chi$, along the direction $\boldsymbol{\d}_i$, to its neighboring site with sub lattice index $\chi_i$ (see Fig.\ref{fig:mft}(A)). The function $\boldsymbol{B}^{\s\s'}_\chi(\o)$ is given by
\be\lb{defB}
\boldsymbol{B}^{\s\s'}_\chi(\o)=\sum_{\d\d'}\frac{\langle L^\chi_{\d\d}\rangle-\langle L^\chi_{\d'\d'}\rangle}{\o+E^\chi_\d-E^\chi_{\d'}}D^{\s\chi}_{\d\d'}D^{\s'\chi}_{\d'\d},
\ee
where the average is taken over the mean field ground state. The excitation spectrum is determined implicitly by setting $\det[\boldsymbol{G}^{-1}(\o,{\bf k})]=0$.

In the Mott regime, there is no superfluid order and the situation simplifies considerably. Eq.(\ref{eomG}) becomes block diagonal with $\boldsymbol{G}^{++}=\boldsymbol{G}^{--}=0$ and we have $(1+\mathcal{B})\boldsymbol{G}^{-+}=\boldsymbol{J}$, where explicitly 
\be
\mathcal{B}=\m{-2J\sin q_y\boldsymbol{B}_A^{-+}}{K(e^{iq_x}-ie^{-iq_x})\boldsymbol{B}_A^{-+}}{K(ie^{iq_x}+e^{-iq_x})\boldsymbol{B}_B^{-+}}{2K\cos q_y\boldsymbol{B}_B^{-+}}
\ee
and the matrix $\boldsymbol{J}$ is given by
\be
\boldsymbol{J}=\m{\boldsymbol{B}_A^{-+}}{0}{0}{\boldsymbol{B}_B^{-+}}.
\ee
As a result, the excitation spectrum is determined by $\det(1+\mathcal{B})=0$ and let us denote it as $E({\bf k})$. There is also a similar branch for the Green function $\boldsymbol{G}^{+-}$, whose solution is given by $E'({\bf k})$. We note the following relation $E({\bf k})=-E'(-{\bf k})$ and will concentrate on $E({\bf k})$ below.

In Fig.\ref{mott_excitation}, we show the two branches of quasi-hole excitations closest to zero energy. One particular feature that is worth noting is the appearance of Dirac points in the excitation spectrum at the positions corresponding to the non-interacting case or their symmetry related points (see Fig.\ref{mott_excitation} (A)). The dispersion close to the Dirac point is linear and can be characterized by two velocities $v_{x,y}$ along $k_{x,y}$, respectively. While the positions of the Dirac points are unaffected by the strong interactions, its dispersion is significantly renormalized from the non-interacting values, as shown In Fig.\ref{mott_excitation} (B). In the deep Mott regime $U\gg J,K$, quasi-particle are essentially doublon and hole~\cite{Chudnovskiy2012}, which hop with a phase relations that is the same as the non-interacting case, apart from the renormalization of the amplitudes and an overall energy shift. As one approaches the Mott-superfluid transition boundary, the Dirac cone becomes shaper. Surprisingly the ``anisotropy" of the Dirac cone, $\frac{v_x}{v_y}=\sqrt{2}$, remains the same as in the non-interacting case. In Fig.\ref{mott_excitation} (C), we plot the quasi-particle and quasi-hole excitation closest to zero energy along a representative path in the first Brillouin zone, connecting symmetry points $\G=(0,0)$, $M=(\frac{\pi}{2},0)$ and $X=(\frac{\pi}{2},\pi)$. As expected, a finite gap always exists in the Mott insulating regime and there is no particle-hole symmetry.

The above features can be understood from the structure of the Green function $\boldsymbol{G}^{-+}$.  Its inverse can be written as: $(\boldsymbol{G}^{-+})^{-1}=\boldsymbol{J}^{-1}(1+\mathcal{B})\equiv h_0+\boldsymbol{h}\cdot\boldsymbol{\s}$. In the case we are considering, namely one boson per site, $\boldsymbol{B}_A^{-+}=\boldsymbol{B}_B^{-+}$ and we find $h_x=H_x$, $h_y=H_y$ and $h_z=H_z$, the same as the non-interacting case. With this, it is straightforward to conclude that the Dirac points will remain at their original positions and in addition, the quasi-particle excitations have the same spinor wave functions as that of a single particle.

Finally, let us discuss the behavior of collective excitations close to the Mott-superfluids transition and show how it is connected to the emergent superfluid states. Fig.\ref{excitation} (A) shows that for fixed ratio $J/K=1$, the excitation energy approaches zero at momentum $(-\frac{\pi}{4},-\frac{\pi}{4})$, which corresponds to the modulations of the plane wave phase. On the other hand, for fixed ratio $J/K=2$ (see Fig.\ref{excitation} (B)), the excitation energy approaches zero at {\em two} distinct {\bf k} points $(-\frac{\pi}{4},-\frac{\pi}{4}+q)$ and $(-\frac{\pi}{4},-\frac{\pi}{4}-q)$, where $q$ depends on the values $J$ and $K$, as well as interaction $U$. This suggests that the emergent condensate is of the form Eq.(\ref{sp}). In Fig.\ref{excitation} (C) and (D), clear difference between the phase modulations in the PP and SP states are shown with overall $\exp(-i\frac{\pi}{4}n)$ factor removed for clear comparison.  


{\it Conclusion}. We have shown how weak coupling superfluid states evolve into the Mott insulating state in a Bose-Hubbard model with synthetic staggered flux. It is predicted that a tricritical point exists where the plane wave state, the stripe state and the Mott insulating state terminate. While excitations in the Mott regime give further supporting evidence of the existence of tricritical point, further studies are necessary to illustrate its nature. Effects of interaction on topological Dirac points are quantified and discussed.



We thank Tin-Lun Ho, Yun Li, Zhenhua Yu and  Hui Zhai for valuable discussions and suggestions. Support from university postgraduate fellowship and postgraduate scholarship (Y.J.) and the startup grant from University of Hong Kong (S.Z.) are gratefully acknowledged.

\bibliography{fluxlattice}

\begin{thebibliography}{29}
\bibitem [{com()}]{com1}%
  \BibitemOpen
  \href@noop {} {\bibinfo  {journal} {Larger cluster sizes with $16$ sites
  along $\hat{x}$ direction and variable number of sites along $\hat{y}$
  direction, from $8$ to $100$ have been used to explore the phases, with
  similar conclusions.}\ }\BibitemShut {NoStop}%
\end{thebibliography}
\end{document}